\documentclass[10pt, conference, compsocconf]{IEEEtran}
\IEEEoverridecommandlockouts
\usepackage[toc,acronym]{glossaries}
\usepackage{adjustbox}
\usepackage{tikz}
\usetikzlibrary{arrows,shadows,backgrounds,calc,decorations.pathreplacing,decorations.pathmorphing,positioning, 
	automata}
\usepackage{pgfplots}
\pgfplotsset{compat=1.13} 
\usepackage{pgfplotstable}
\makeglossaries

\newacronym{FPGA}{FPGA}{Field Programmable Gate Array}  
\newacronym{HW}{HW}{hardware}
\newacronym{ISA}{ISA}{Instruction Set Architecture}
\newacronym{I/O}{I/O}{Input/Output}
\newacronym{MC}{MC}{Multi-Core and/or Many-Core}
\newacronym{MLP}{MLP}{Memory Level Parallelism}
\newacronym{OoO}{OoO}{Out-of-Order}
\newacronym{OS}{OS}{operating system}
\newacronym{PD}{PD}{Propagation Delay}
\newacronym{PU}{PU}{Processing Unit}
\newacronym{SPA}{SPA}{Single Processor Approach}
\newacronym{SW}{SW}{software}
\newacronym{HPL}{HPL}{High Performance Linpack}
\newacronym{HPCG}{HPCG}{High Performance Conjugate Gradients}

\pgfplotscreateplotcyclelist{my color list}{%
	solid, color=webblue, every mark/.append style={solid, fill=webblue}, mark=*\\%
	densely dashdotted, color=webred, every mark/.append style={solid, fill=webred},mark=diamond*\\%
	densely dotted, color=webgreen, every mark/.append style={solid, fill=webgreen}, mark=triangle*\\%
	loosely dashed, color=webbrown, every mark/.append style={solid, fill=webbrown},mark=square,fill=webgreen!20\\%
	densely dotted, every mark/.append style={solid, fill=gray}, mark=otimes*\\%
	dashed, every mark/.append style={solid, fill=gray},mark=diamond*\\%
	densely dashed, every mark/.append style={solid, fill=gray},mark=square*\\%
	dashdotted, every mark/.append style={solid, fill=gray},mark=otimes*\\%
	dashdotdotted, every mark/.append style={solid},mark=star\\%
}
\definecolor{webgreen}{rgb}{0,.5,0}
\definecolor{webbrown}{rgb}{.6,0,0}
\definecolor{webyellow}{rgb}{0.98,0.92,0.73}
\definecolor{webgray}{rgb}{.753,.753,.753}
\definecolor{webblue}{rgb}{0,0,.8}
\definecolor{webgreen}{rgb}{0, 0.5, 0} % less intense green
\definecolor{webred}{rgb}{0.5, 0, 0}   % less intense red

\usepackage{cite}
\usepackage{amsmath,amssymb,amsfonts}
\usepackage{graphicx}
\usepackage{textcomp}
\usepackage{xcolor}
\def\BibTeX{{\rm B\kern-.05em{\sc i\kern-.025em b}\kern-.08em
    T\kern-.1667em\lower.7ex\hbox{E}\kern-.125emX}}

\def\LightSpeed{3.e8}	% m/s
\def\Gravity{10.}		%m/s^2
\def\Speed{x*\Gravity}
\def\Density{1.}
\def\RelSpeedFactor{\Speed/(\LightSpeed/\Density)}
\def\RelSpeed{\Speed/sqrt(1+\RelSpeedFactor*\RelSpeedFactor)}

\def\RelSpeedFactorB{\Speed/(\LightSpeed/\Density/2)}
\def\RelSpeedB{\Speed/sqrt(1+\RelSpeedFactorB*\RelSpeedFactorB)}

\def\OneDay{86400}

\begin{document}

\title{The need for modern computing paradigm:\\
	Science applied to computing\\
\thanks{Project no. 125547  has been implemented with the support provided from the National Research, Development and Innovation Fund of Hungary, financed under the K funding scheme.
		Also the ERC-ECAS support of project 886183 is acknowledged.}
}

\author{\IEEEauthorblockN{J\'anos  V\'egh}
\IEEEauthorblockA{\textit{Kalim\'anos BT} \\
Debrecen, Hungary \\
Vegh.Janos@gmail.com~ORCID:~0000-0002-3247-7810}
\and
\IEEEauthorblockN{Alin Tisan}
\IEEEauthorblockA{\textit{Royal Holloway}, University of London, UK \\
	\textit{Department of Electronic Engineering}\\
 alin.tisan@rhul.ac.uk~ORCID:~0000-0002-8280-2722 }
}

\maketitle

\begin{abstract}
More than hundred years ago the ’classic physics’ was in its full power, with just a few unexplained phenomena; which, however, led to a revolution and the development of the ’modern physics’. The outbreak was possible by studying the nature under extreme conditions which finally led to the understanding of the relativistic and quantal behavior. Today, the computing is in a similar position: it is a sound success story, with exponentially growing utilization but, as moving towards extreme utilization conditions, with a growing number of difficulties and unexpected issues which cannot be explained based on the ’classic computing paradigm’. The paper draws the attention that under extreme conditions, computing behavior could differ than the one in the normal conditions, and pinpoints that certain, unnoticed or neglected features enable the explanation of the new phenomena, and the enhancement of some computing features. Moreover, a new modern computing paradigm implementation idea is proposed.
\end{abstract}


%
%Type: Full/Regular Research Papers
%
%Relevant symposium: Parallel \& Distributed Computing (CSCI-ISPD)
%
%\pagebreak
%% main text
\section{Introduction}\label{sec:Introduction}

Initially computers were constructed with the goal to reduce the computation time of rather complex but sequential mathematical tasks~\cite{EDVACEckertMauchly,EDVACreport1945,FateofEDVAC1993,
	GodfreyIEEE1993}.
% (e.g. the computation of trajectories for missiles).
%Therefore their architecture was designed to enable that kind of activity, and even it was recognized early at that time that that architecture is not really efficient in solving certain kinds of problems.
Today a computer is deployed for radically different goals, where the mathematical computation in most cases is just a very small part of the task. The majority of the non-computational activities are reduced to or imitated by some kind of \textit{computations, because this is the only activity that the \textit{computer} can do}.

Because of this, the current computer architecture (concluded from the 70-year old 'classic' paradigm)
is not really suitable for the goals they are expected to work for:
to handle the varying degree of parallelism in a dynamic way;
to work in multi-tasking mode;
to prepare systems comprising many-many  processors with a good efficacy;
to cooperate with the large number of other processors (both on-chip and off-chip);
to monitor many external actions and react to them in real-time;
%--to build servers in environments where the workload changes in an extreme way;
to operate 'big data' processing systems, etc.
%--Under the conditions the computers are utilized today
Not only the efficacy of computing is very low because of the different performance 
losses~\cite{InefficiencyHameed2010}, but more and more limitations come to the light~\cite{NeuromorphicComputing:2015}.
%--,ChandyParallelism:2009}. 

Not only that in extreme conditions computing quickly reaches  its performance limits~\cite{NeuromorphicComputing:2015,LimitsOfLimits2014},
but this performance degrades even more accelerated in systems comprising parallelized sequential processors~\cite{VeghPerformanceWall:2019}.
The very rigid HW-SW separation leads to the phenomenon known as ’priority inversion’~\cite{PriorityInversion:1993},
the very large rigid architectures suffer from frequent component errors,
the complex cyber-physical systems
must be equipped with excessive computing facilities to provide real-time operation,
delivering large amount of data from the big storage centres to the places where the processing capacity
is concentrated consitutes a major challenge, etc.
All these issues have a  common reason~\cite{SoOS:2010}: the classical computing paradigm that reflects the state of the art
of computing 70 years ago. Computing needs renewal~\cite{RenewingComputingVegh:2018}. 
Not only the materials and methods of gate handling~\cite{RebootingComputingModels:2019} but also our thinking must be rebooted.

It was early discovered that the age of conventional architectures was over ~\cite{AmdahlSingleProcessor67,GodfreyArchitecture1986};
 the only question that remained open whether the “game is over”, too~\cite{ComputingPerformanceBook:2011}. 
 When comes to parallel computing, the today’s technology is able to deliver not only many, but "too many" cores~\cite{TooManyCores2007}.
Their computing performance, however, does not increase linearly with the number of cores~\cite{HillMulticoreAmdahl2008} 
("\textit{a trend that can't go on ad infinitum}"\cite{ExascaleGrandfatherHPC:2019}),
and only a few of them can be utilized simultaneously ~\cite{Computing_Dark_Silicon_2017}.  
Consequently, it is unreasonable to use high number of cores for such extreme system operation~\cite{Ungerer_MERASA:2010}.
Approaching the limits of the ”classic paradigm of computing” has rearranged the technology ranking~\cite{KeepingComputerIndustryInUS:2017}.
"\textit{New ways of exploiting the silicon real estate need to be explored}"~\cite{ChandyParallelism:2009} and a modern computing paradigm is needed indeed.

One of the major issues is,  that "\textit{%Development of software is expensive and requires a
%	highly educated workforce. 
	being able to run software
	developed for today's computers on tomorrow's
	has reigned in the cost of software development.
	But this guarantee between computer designers
	and software designers has become a barrier to a
	new computing era.}"~\cite{KeepingComputerIndustryInUS:2017}
A new computer paradigm, that is an extension rather than a replacement of the old paradigm,
has better chances to be accepted, since it provides a smooth transition
from the age of old computing to the age of modern computing.

\section{Analogies with the classic versus modern physics}

The case of computing is very much analogous with the case of classic physics versus the modern physics (relativistic and quantum). 
Given the world we live in it is rather counter-intuitive to accept
that  as we move towards unusual conditions, the adding of speeds behaves differently, 
%when approaching the speed of light as well as that
the energy becomes discontinuous, the momentum and the position of a particle cannot be measured accurately at the same time.
%Basically, these are curiosities or nuances for those who deploy their knowledge in physics in the everyday life, and 
In normal cases there is no notable difference, when using if applying or not the non-classical principles.
However, as we get farther from the everyday  conditions, the difference gets more considerable, and even leads to phenomena one can never experience under the usual, everyday conditions. \textit{The analogies do not want to imply 
direct correspondence between certain physical  and 
computing phenomena. Rather, the paper draws the attention to both
that under extreme conditions qualitatively different behavior
may be encountered, and that scrutinizing certain, formerly unnoticed or neglected  aspects enables to explain the new phenomena. %In computing, u
Unlike in the nature, the technical implementation of the critical points can be changed, and through this
the behavior of the computing systems can also be changed.}
In this paper, only two of the affected important areas of computing  will be touched in more details: parallel processing and multitasking.

\subsection{Analogy with the special relativity}
\label{sec:AnalogyRelativity}

%--In the above sense, 
In the above sense, there is an important difference between the operation and the performance of the single-processor and those of the parallelized but sequentially working computer systems.
As long as just a few (thousands) single processors are aggregated into a large computer system, the resulting performance will correspond (approximately) to the sum of the single-processor performance values: similarly to the classic rule of adding speeds (see also table~\ref{tab:summing}).
However, when assembling larger computing systems (and approaching with their performance "the speed of light"
of computing systems in the range of millions of processors) the \textit{nominal} performance
starts to deviate from the experienced \textit{payload} performance:
the phenomenon known as \textit{efficiency} appears. In addition, there is more than just \textit{one} efficiency~\cite{DifferentBenchmarks:2017}:
\textit{the measurable efficiency depends on the method of measurement} (the utilized benchmark program). Actually, the simple Amdahl's Law results in a 2-dim surface\footnote{As can be seen from Figure~6, at extreme high number of cores  $(1-\alpha)$ also depends on the number of cores.} shown in Figure~1. Changing the implementation method, the efficacy at large number of cores can be an order of magnitude higher.

\begin{table}
	\begin{tabular}{|p{75pt}|p{150pt}|}
		\hline
		\hline
		Physics & Computing\\
		\hline
		Adding of speeds &	Adding of performance\\
		\hline
		\textcolor{blue}{Classic} & \textcolor{blue}{Classic} \\
		\textcolor{blue}{$ v(t) = t\times a$}
		&
		\textcolor{blue}{$ Perf_{total}(N) = N\times Perf_{single}$}	\\
		\hline
		
		c = Light Speed &\\
		\hline
		t = time &  N = number of cores\\
		\hline
		a = acceleration & $Perf_{single}$ = single processor performance\\
		\hline
		n = optical density
		&
		$\alpha$ = parallelism\\
		\hline
		\textcolor{red}{ Modern (relativistic)} &\textcolor{red}{Modern\cite{VeghAlphaEff:2016}} \\
		\textcolor{red}{$ v(t) = \frac{t\times a}{\sqrt{1+{(\frac{t\times g}{c/n}})^2}}$}
		&
		\textcolor{red}{$ Perf_{total}(N) = \frac{N\times Perf_{single}}{n\times (1-\alpha)+\alpha}$}	
		\\
		\hline
		\hline
		
	\end{tabular}
\vskip\baselineskip
	\caption{The "adding speeds" analogies between the classic and modern arts of science and computing, respectively}
	\label{tab:summing}
\vskip-2\baselineskip
	
\end{table}

\begin{figure}
\vskip-\baselineskip
\hskip-.7cm	\maxsizebox{1.2\columnwidth}{!}
	{
		\includegraphics{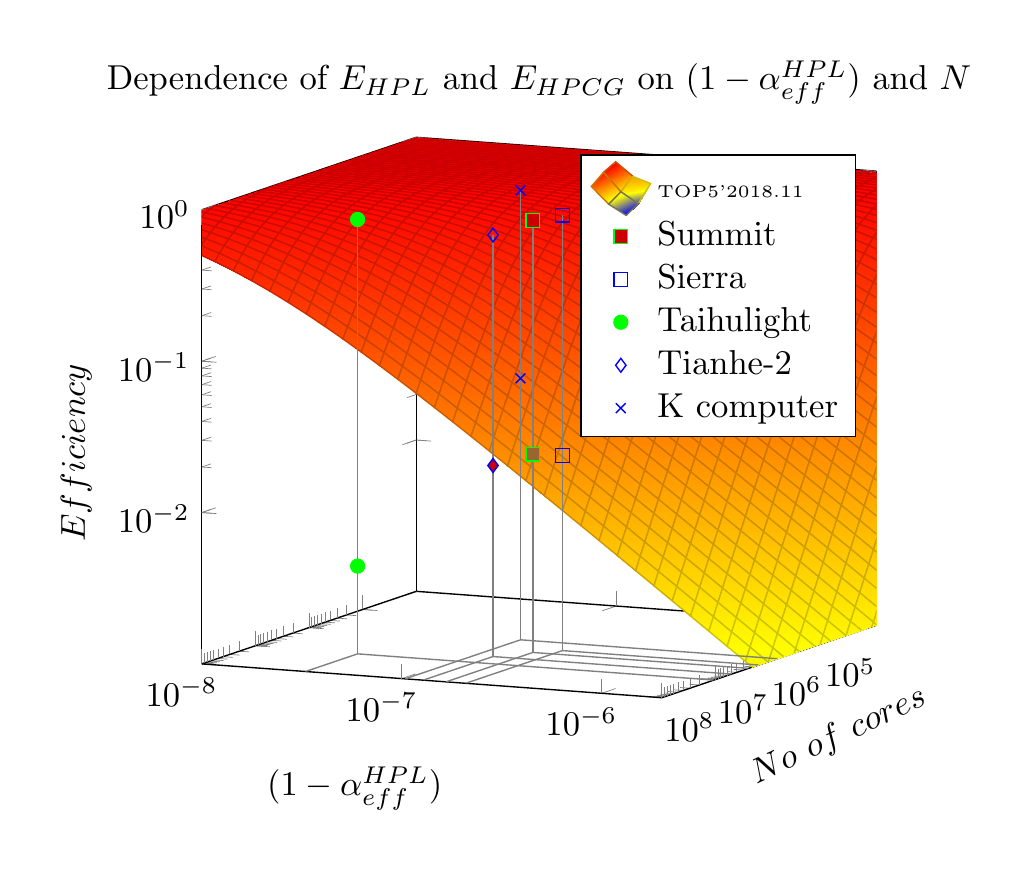}
	}
\vskip-\baselineskip
	\label{fig:Efficiency surface}
	\caption{
		The efficiency surface corresponding 
	to the "modern paradigm", see Table~\ref{tab:summing}, with some measured (with $HPL$ and $HPCG$) efficiencies of supercomputers. Recall that "\textit{this decay in performance is not a fault of the architecture, but is dictated by the limited parallelism}".\cite{ScalingParallel:1993}
}
\vskip-\baselineskip
\end{figure}

\begin{figure*}
	\maxsizebox{\textwidth}{!}
	{\includegraphics{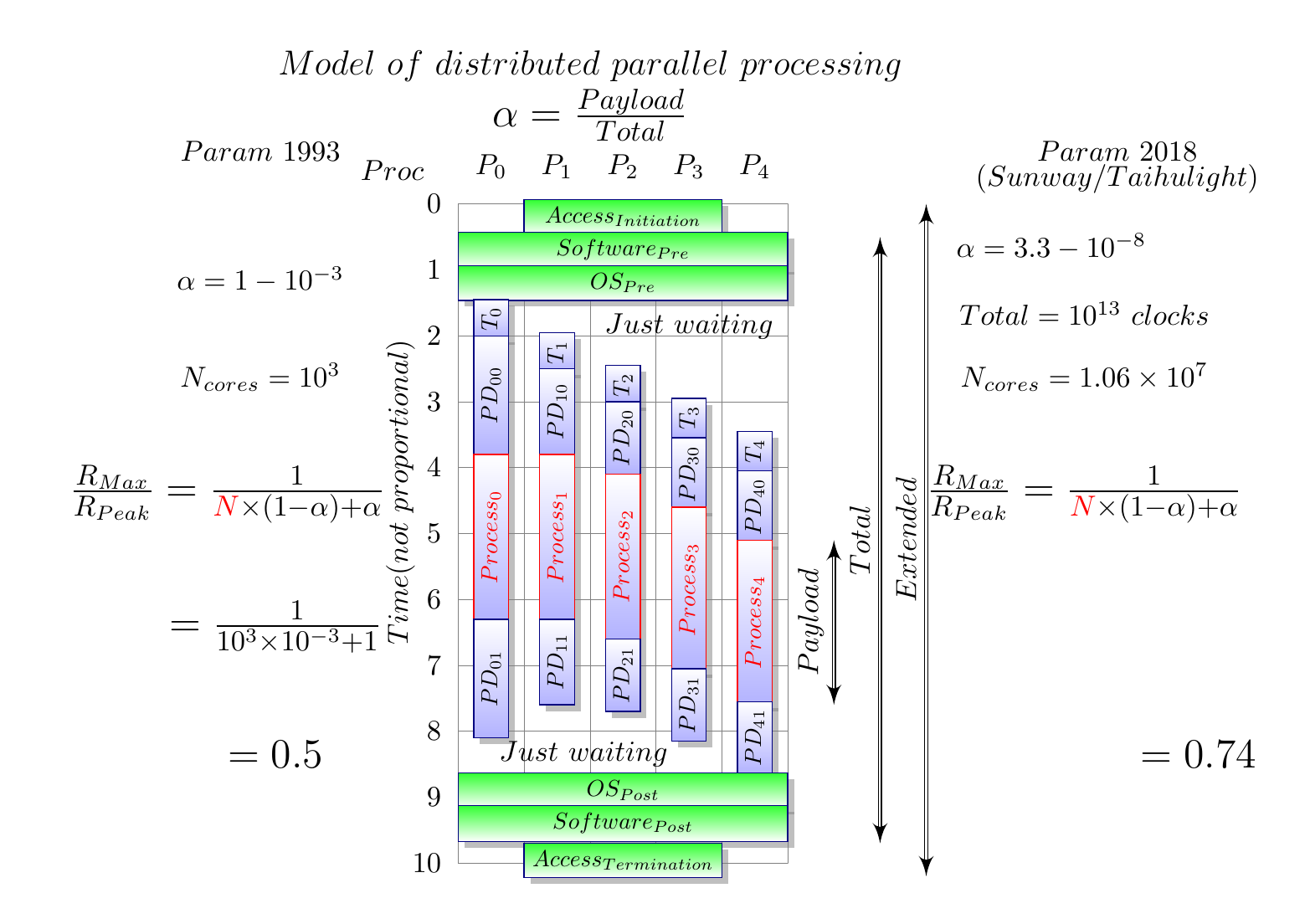}
	}
\vskip-\baselineskip
	\label{fig:AmdahlModel}
	\caption{A general model of parallel operation. For better 
		visibility, the lengths of the boxes are not proportional with the time the corresponding action needs. The data are taken from~\cite{ScalingParallel:1993} and ~\cite{DongarraSunwaySystem:2016}}
\vskip-\baselineskip
\end{figure*}

What makes the analogy really relevant, it is shown in Table~\ref{tab:summing}: although the function form
is different, the modern approach to computing (see the model) introduces a 
correction term that remains close to unity 
until the extreme large number of cores is approached, then
the performance functions saturate.
The phenomenon is surprisingly similar to increasing the speed of a body with constant acceleration (see Figure~5%\ref{fig:lightspeed}
): at low time (or speed) values 
there is no noticeable difference between the speeds calculated with or without relativistic correction,
but as speed closes to the speed of light, the speed of the accelerated body saturates. Whether it is formulated that the mass increases or the time slows down, the essence is the same: when approaching (the specific) limits, the behavior of the subject drastically changes.

The performance measurements are simple time measurements: a standardized set of machine instructions is executed
(a large number of times) and the known number of operations is divided by the measurement time;
both for the single-processor and for the distributed parallelized sequential systems.
%--This happens in the same way in the case of measuring the performance
%--of both the single-processor and the parallelized sequential computing systems.
In the latter case, however, the joint work must also be organized, %--i.e. an extra task (
implemented with extra machine instructions and extra execution time. %--) appears.
As Fig.~2 shows, in the parallel operating mode \textbf{\textit{both the software
and the 
hardware contribute to the execution time}}, i.e. both must be considered in Amdahl's Law.
%, and the actions
%are partly parallel, partly sequential.

%One of the processors must tell the others (at least) what fraction they should perform and also
%the result of the calculations must be collected.
%This organization activity is needed, but it takes (non-payload) time from the point of view of the calculation.

This is the origin of the efficiency: one of the processors orchestrates the joint operation, the others are waiting.
%; all of them 
%wasting a fraction of the measured operating time to
%% (from the point of view of calculation) 
%non-payload activity.
%%Since the first processor must speak to the others one-by-one, the wasted time increases linearly with the number of concerted processors.
After reaching a certain number of processors there is no more increase in the payload fraction when adding more processors:
the first fellow processor already finished the task and is waiting while the last one is still waiting for the start command.
%This limiting number can be increased by organizing the processors into clusters:
%then the first computer must speak \textit{directly} only to the head of the cluster.
The physical size of the computing system also matters:
the processor connected with a cable of length of dozens of meters to the first one  must spend several hundreds clock cycles with waiting (not mentioning geographically distributed computer systems, such as some clouds, connected through general-purpose networks).
 Detailed calculations are given in~\cite{Vegh:2017:AlphaEff}.
 
%When combining properly the propagation delay (PD) with
%the sequential scheduling, the non-payload time can be considerably reduced during fine-tuning the system
%(in the case of Sierra 0\% increase in the number of cores
%resulted in 32\% more performance, in the case Summit
%a 5\% increase in the number of cores resulted in 17\%
%more performance).
%Also, mismatching the total time and the extended measurement time may lead to completely wrong conclusions~\cite{BenchmarkingClouds:2017} as discussed in~\cite{Vegh:2017:AlphaEff}.

The phenomenon itself is known for decades
\cite{ScalingParallel:1993}:
"\textit{Amdahl argued that most parallel programs have
	some portion of their execution that is inherently serial
	and must be executed by a single processor while others
	remain idle. \dots In fact, \textbf{there comes a point when using more processors \dots actually increases the execution time rather than reducing it}.}"
Presently however the theory was almost forgotten mainly due to the quick development of the parallelization technology and the increase of the single-processor performance.
%, as well as the considerably longer benchmarking time.

During the past quarter of century, the proportion of the contributions changed considerably: 
today the number of processors is thousands of 
times higher than a quarter of century ago, 
the growing physical size and the higher processing speed
increased the role of the propagation delay.  
As a result of the technical development the same phenomenon returned in a technically 
different form at much higher number of processors.
Due to this, \textit{the computing performance cannot be increased above the performance defined by the parallelization technology and the number of processors} (this is similar to affirming that an object having the speed of light cannot be further accelerated). Exceeding a certain computing performance (using the classic paradigm and its classic implementation) is prohibited by the laws of nature. 

%Despite of the huge competition, the payload performance changes with relatively small value as the time passes.
The "speed of light" limit is specific for the different architectures, but the higher is the achieved performance, the harder is to increase it.
%Considerable performance increase happened only at relatively low
%computing performance values. 
It looks like that in the feasibility studies  an analysis
whether some inherent performance bound exists remained out of sight either in USA~\cite{NSA_DOE_HPC_Report_2016,Scienceexascale:2018} or in  EU~\cite{EUActionPlan:2016}
or in Japan~\cite{JapanExascale:2018} or in China~\cite{ChinaExascale:2018}.

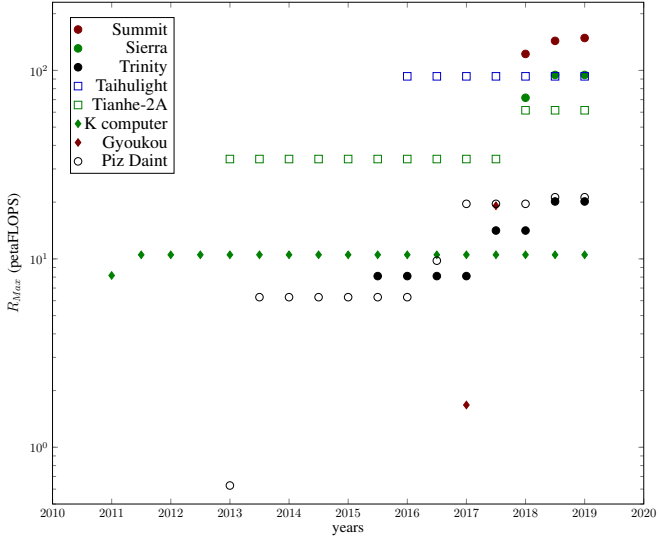
\begin{figure}
	\begin{center}
		\maxsizebox{\columnwidth}{!}{
			\begin{tikzpicture}[scale=0.95]
			\begin{axis}
			[
			width=\textwidth,
			cycle list name={my color list},
			legend style={
				cells={anchor=east},
				legend pos={north west},
			},
			xmin=2010, xmax=2020,% x scale
			ymin=.5e0, ymax=230, % y scale
			xlabel={\large years},
			/pgf/number format/1000 sep={},
			ylabel={\large $R_{Max}$  (petaFLOPS)},
			%		xmode=log,
			%		log basis x=10,
			ymode=log,
			log basis y=10,
			]
			\addplot[only marks,  mark=*,  color=webred, mark size=3, thick] plot coordinates {
				(2018.0,122.3) %06/2018
				(2018.5,143.5) %11/2018
				(2019.0,148.6) %06/2018
			};
			\addlegendentry{\Large Summit}
			\addplot[only marks,  mark=*,  color=webgreen, mark size=3, thick] plot coordinates {
				(2018.0,71.6) %06/2018
				(2018.5,94.6) %11/2018
				(2019.0,94.6) %06/2018
			};
			\addlegendentry{\Large Sierra}
			\addplot[only marks,  mark=*,  mark size=3, thick] plot coordinates {
				(2015.5,8.101) %11/2015
				(2016.0,8.101) %06/2016
				(2016.5,8.101) %11/2016
				(2017.0,8.101) %06/2017
				(2017.5,14.137) %11/2017
				(2018.0,14.137) %06/2018
				(2018.5,20.159) %11/2018
				(2019.0,20.159) %06/2018
			};
			\addlegendentry{\Large Trinity}
			\addplot[only marks,  mark=square,  color=webblue, mark size=3, thick] plot coordinates {
				(2016.0,93.015) %06/2016
				(2016.5,93.015) %11/2016
				(2017.0,93.015) %06/2017
				(2017.5,93.015) %11/2017
				(2018.0,93.015) %06/2018
				(2018.5,93.015) %11/2018
				(2019.0,93.015) %06/2019
			};
			\addlegendentry{\Large Taihulight}
			\addplot[only marks,  mark=square,  color=webgreen, mark size=3, thick] plot coordinates {
				(2013.0,33.863) %06/2013
				(2013.5,33.863) %11/2013
				(2014.0,33.863) %06/2014
				(2014.5,33.863) %11/2014
				(2015.0,33.863) %06/2015
				(2015.5,33.863) %11/2015
				(2016.0,33.863) %06/2016
				(2016.5,33.863) %11/2016
				(2017.0,33.863) %06/2017
				(2017.5,33.863) %11/2017
				(2018.0,61.445) %06/2018
				(2018.5,61.445) %11/2018
				(2019.0,61.445) %06/2019
			};
			\addlegendentry{\Large Tianhe-2A}
			\addplot[only marks,  mark=diamond*,  color=webgreen, mark size=3, thick] plot coordinates {
				(2011.0,8.162) %06/2011
				(2011.5,10.51) %11/2013
				(2012.0,10.51) %06/2014
				(2012.5,10.51) %11/2012
				(2013.0,10.51) %06/2013
				(2013.5,10.51) %11/2013
				(2014.0,10.51) %06/2014
				(2014.5,10.51) %11/2014
				(2015.0,10.51) %06/2015
				(2015.5,10.51) %11/2015
				(2016.0,10.51) %06/2016
				(2016.5,10.51) %11/2016
				(2017.0,10.51) %06/2017
				(2017.5,10.51) %11/2017
				(2018.0,10.51) %06/2018
				(2018.5,10.51) %11/2018
				(2019.0,10.51) %06/2019
			};
			\addlegendentry{\Large K computer}
			\addplot[only marks,  mark=diamond*,  color=webred, mark size=3, thick] plot coordinates {
				(2017.5,19.136) %11/2017
				(2017.0,1.677) %06/2017
			};
			\addlegendentry{\Large Gyoukou}
			\addplot[only marks,  mark=o,  mark size=3, thick] plot coordinates {
				(2012.5,0.2164) %11/2012
				(2013.0,0.6269) %06/3013
				(2013.5,6.261) %11/2013
				(2014.0,6.261) %06/2014
				(2014.5,6.261) %11/2014
				(2015.0,6.261) %06/2015
				(2015.5,6.261) %11/2015
				(2016.0,6.261) %06/2016
				(2016.5,9.779) %11/2016
				(2017.0,19.590) %06/2017
				(2017.5,19.590) %11/2017
				(2018.0,19.590) %06/2018
				(2018.5,21.230) %11/2018
				(2019.0,21.23) %06/2018
			};
			\addlegendentry{\Large Piz Daint}
			\end{axis}
			\end{tikzpicture}
		}
		\caption{The $R_{Max}$ payload performance in function of the year of construction for different configurations. The performances of all configurations seem to have their individual saturation value. The later a supercomputer appears
			in the competition, the smaller is the performance 	ratio with respect 
			to its predecessor; the higher is its rank, the harder is to improve its performance.
		}
		\label{fig:SupercompSpeedOfLight}
	\end{center}
\vskip-\baselineskip
\end{figure}

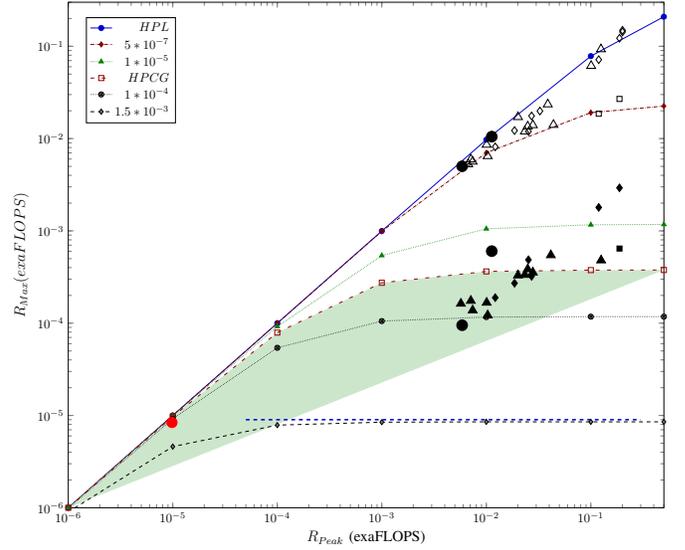
\begin{figure}
	\begin{center}
		\maxsizebox{\columnwidth}{!}{
			\begin{tikzpicture}[scale=0.95]
			\begin{axis}
			[
			width=\textwidth,
			cycle list name={my color list},
			legend style={
				cells={anchor=east},
				legend pos={north west},
			},
			xmin=1e-6, xmax=0.5,% x scale
			ymin=1e-6, ymax=0.3, % y scale
			xlabel={\large $R_{Peak}$  (exaFLOPS)},
			/pgf/number format/1000 sep={},
			ylabel={\large $R_{Max} (exaFLOPS)$},
			xmode=log,
			log basis x=10,
			ymode=log,
			log basis y=10,
			]
			%\addplot table [x=a, y=i, col sep=comma] {RMaxvsRPeakatAlpha.csv};
			%		\addlegendentry{$1*10^{-8}$}
			\addplot table [x=a, y=h, col sep=comma] {RMaxvsRPeakatAlpha.csv};
			\addlegendentry{$HPL$}
			%\addplot table [x=a, y=g, col sep=comma] {RMaxvsRPeakatAlpha.csv};
			%		\addlegendentry{$1*10^{-7}$}
			\addplot table [x=a, y=k, col sep=comma] {RMaxvsRPeakatAlpha.csv};
			\addlegendentry{$5*10^{-7}$}
			%\addplot table [x=a, y=f, col sep=comma] {RMaxvsRPeakatAlpha.csv};
			%		\addlegendentry{$1*10^{-6}$}
			\addplot table [x=a, y=e, col sep=comma] {RMaxvsRPeakatAlpha.csv};
			\addlegendentry{$1*10^{-5}$}
			\addplot table [x=a, y=d, col sep=comma] {RMaxvsRPeakatAlpha.csv};
			\addlegendentry{$HPCG$}
			\addplot table [x=a, y=c, col sep=comma] {RMaxvsRPeakatAlpha.csv};
			\addlegendentry{$1*10^{-4}$}
			%\addplot table [x=a, y=b, col sep=comma] {RMaxvsRPeakatAlpha.csv};
			%		\addlegendentry{$3*10^{-4}$}
			\addplot table [x=a, y=l, col sep=comma] {RMaxvsRPeakatAlpha.csv};
			\addlegendentry{$1.5*10^{-3}$}
			% Supercomputers TOP10 by HPCG@2018 June
			% Non-accelerated; position in Jun2018 (HPCG, HPL)
			\addplot[only marks,  mark=triangle,  mark size=4, thick] plot coordinates {
				(0.01127,0.0105) %K (3,16) computer at HPL
				(0.0439,0.0141) %Trinity (4,9) at HPL
				(0.1007, 0.0614) %Tianhe-2A (4,?) at HPL
				(0.125,0.0930) %Sunway (6,2) at HPL
				(0.0249,  0.013554) %Oakforest-PACS (7,12) at HPL
				(0.0279,  0.0140) %Cori (8,10) at HPL
				(0.0234,  0.01197) %Tera-1000-2 (9,14) at HPL
				(0.0201,  0.01713) %Sequoia (10,8) at HPL
				(0.00711, 0.00595) %Pleiades (14,24) at HPL
				(0.01007, 0.00859) %Mira (15,17) at HPL
				(0.00671, 0.00528) %Pangea (16,30) at HPL
				(0.00740, 0.00564) %Hazel Hen (17,27) at HPL
				(0.01030, 0.00647) %MareNostrum (18,22) at HPL
				(0.03875, 0.02352) % Frontera (5,?) at HPL
			};
			\addplot[only marks,  mark=triangle*,  mark size=4, thick] plot coordinates {
				(0.01128,0.000602) %K computer at HPCG
				(0.04146,0.000546) %Trinity at HPCG
				(0.1254, 0.000480) %Sunway at HPCG
				(0.0249, 0.000385) %Oakforest-PACS at HPCg
				(0.0279, 0.000355) %Cori at HPCg
				(0.0234, 0.000334) %Tera-1000-2 at HPCg
				(0.0201, 0.000330) %Sequoia  at HPCg
				(0.00711,0.000175) %Pleiades at HPCG
				(0.01007,0.000167) %Mira at HPCG
				(0.00571,0.000163) %Pangea at HPCG
				(0.00740,0.000138) %Hazel Hen at HPCG
				(0.01030,0.000122) %MareNostrum at HPCG
				
			};
			% Nvidia accelerated
			\addplot[only marks,  mark=diamond,  mark size=3, thick] plot coordinates {
				(0.188,0.1223) %Summit (1,1) at HPL (06/2018)
				(0.201,0.143) %Summit (1,1) at HPL (11/2018)
				(0.201,0.149) %Summit (1,1) at HPL (06/2019)
				(0.119,0.0716) %Sierra (2,3) at HPL
				(0.03258, 0.01988) %AI Bridging Cloud (5,?) at HPL
				(0.0253,0.01196) %Piz Daint (5,6) at HPL
				(0.02711, 0.01759) %Titan (11,7) at HPL
				(0.01862, 0.01221) %HPC4 (12,13) at HPL
				(0.01217, 0.00813) %TSUBAME3.0 (13,19) at HPL
			};
			\addplot[only marks,  mark=square,  mark size=2, thick] plot coordinates {
				(0.188,0.0269) %Summit (1,1) at HPL, corrected
				(0.119,0.0186) %Sierra (2,3) at HPL, corrected
			};
			\addplot[only marks,  mark=diamond*,  mark size=3, thick] plot coordinates {
				(0.188,  0.002926) %Summit at HPCG
				(0.119,  0.001795) %Sierra at HPCG
				(0.0253, 0.000486) %Piz Daint at HPCG
				(0.02711,0.000322) %Titan  at HPCG
				(0.01862,0.000271) %HPC4  at HPCG
				(0.01217,0.000189) %TSUBAME3.0 at HPCG
			};
			\addplot[only marks,  mark=square*,  mark size=2, thick] plot coordinates {
				(0.188,  0.000644) %Summit at HPCG, corrected  (06/2018)
				%			(0.188,  0.000644) %Summit at HPCG, corrected  (11/2018)
				%			(0.119,  0.000466) %Sierra at HPCG, corrected  (06/2019)
			};
			\addplot[only marks,  mark=*,  mark size=4, very thick] plot coordinates {
				(0.01127,0.0105) %K (3,16) computer at HPL
				(0.00587,0.00501) %JUQUEEN computer at HPL
				(0.01127,0.000603) %K (3,16) computer at HPL
				(0.00587,0.000095) %JUQUEEN computer at HPCG
			};
			\addplot[only marks,  mark=*,  mark size=4, color=red, thick] plot coordinates {
				(9.83e-6,8.39e-6) %JUQUEEN computer at Albada
			};
			\draw[very thick,webblue,dashed] (5e-5,9e-06) -- (.3,9e-06);
			\end{axis}
			\end{tikzpicture}
		}
		\caption{The payload performance  $R_{Max}$ in function of the nominal performance $R_{Peak}$, at different $(1-\alpha_{eff})$ values. The figures 
			display the measured values
			derived using \gls{HPL} (empty marks) and \gls{HPCG} (filled marks) benchmarks, for the TOP15 supercomputers (as of 2019 June). The diagram lines marked as  \gls{HPL} and \gls{HPCG} correspond to the behavior of supercomputer $Taihulight$ at $(1-\alpha_{eff})$ values $3.3*10^{-8}$ and $2.4*10^{-5}$, respectively.
			The uncorrected values of the new supercomputers
			$Summit$ and $Sierra$ are shown as diamonds,
			and the same values corrected for single-processor performance are shown as rectangles. 
			The black dots mark the performance data of 
			supercomputers $JUQUEEN$ and $K$ as of 2014 June,
			for \gls{HPL} and \gls{HPCG} benchmarks, respectively.
			The saturation effect can be observed for both
			\gls{HPL} and \gls{HPCG} benchmarks.
			The shaded area only highlights the nonlinearity.
			The red dot denotes the performance value of the system used by \cite{NeuralNetworkPerformance:2018}.
			The dashed line shows the plausible saturation performance value of the brain simulation.
			% The computing performance of AI applications may be between the diagram line marked by \gls{HPCG} and that of the brain simulation. 
		}
		\label{fig:exaRMaxAlpha}
	\end{center}
\vskip-\baselineskip
\end{figure}

%As discussed above (and in details in~\cite{VeghPerformanceWall:2019}),
%the nonlinearity parallelized performance is not noticeable at low performance values and its exact form depends also on the benchmarking task. 
In Fig.~\ref{fig:SupercompSpeedOfLight}, the development of the payload performance of some top supercomputers in function of the year of construction is depicted. 
In Fig.~\ref{fig:exaRMaxAlpha} \textit{the dependence
of payload performance on nominal performance} is
 depicted 
for different benchmark types, specifically for the commonly
utilized \gls{HPL} and \gls{HPCG} benchmarks.
All diagram lines clearly show the signs of saturation~\cite{ExponentialLawsComputing:2017} (or rooflines~\cite{VeghRoofline:2019}): the more inter-processor communication 
is needed for solving the task, the more "dense" is the environment and the smaller is the specific "speed of light". 
The third "roofline" is rather guessed~\cite{VeghBrainAmdahl:2019} for the case of processor-based brain simulation case, where higher orders of magnitude
of communication is required. The payload performance of the \textit{processor-based} Artificial Intelligence tasks, given the amount of communication involved, is expected to be between the latter two rooflines, closer to that of the \gls{HPCG} level.

Supercomputers are computing systems stretched to the limit~\cite{StrechingSupercomputers:2017,Scienceexascale:2018}.
They are a reflection of the engineering perfectness which, in the case of Summit supercomputer,
showed an increase of 17 \% in the computing performance when adding 5 \% more cores (and making fine-tuning after its quick startup)
in its first half year after its appearance, and another  3.5 \% increase in its performance when adding 0.7 \% more cores
after another half a year.
The one-time appearance of Gyoukou is mystic: it could catch slot \#4 on the list using just 12\% of its 20M cores (and was withdrawn immediately)
although its explicit ambition was to be the \#1.
This example shows that the lack of understanding of the computing systems behaviour under extreme conditions
led to the false presumptions of frauds when reporting computing performances ~\cite{GyoukouFraud:2017}.
Another champion candidate, supercomputer Aurora~\cite{DOEAuroraMistery:2017} --after years of building--
was retargeted just weeks before its planned startup. In November 2017   Intel announced
that \textit{Aurora} has been shifted to 2021. As part of the announcement \textit{Knights Hill} was canceled and instead be replaced by a "new platform and new microarchitecture specifically designed for exascale". The lesson learned was that \textit{specific design is needed for exascale}.

As Figure~\ref{fig:exaRMaxAlpha} shows (and discussed in details in~\cite{VeghPerformanceWall:2019}),  the deviation
from the linear dependence is unnoticeable at low performance values: here the "classic speed addition" is valid. At extremely large performance values, 
however, the dependence is strongly non-linear and specific
 on the measurement conditions: here the performance limits manifest. The large scatter of the measured data is generated by the variance of different 
processor and connection types, design ideas, manufacturers, etc.
However, the diagram perfectly reflects the theory that describes the tendency.

\subsection{Analogy with the general relativity}
\label{sec:AnalogyGeneralRelativity}

The mentioned decrease of the system's performance manifests in the appearance of the performance wall~\cite{VeghPerformanceWall:2019,VeghRoofline:2019},
a new limitation due to the parallelized sequential computing.
To make a parallel with the modern physics,
it is known that the objects with extreme large masses 
behave differently from what we know they should behave under 'normal' conditions. 
In other words, the  behavior of large scale 'matter' largely deviates from that
of the small scale 'matter'. This analogy to physics is  the 'dark silicon'~\cite{Computing_Dark_Silicon_2017} whose behaviour resembles the 'dark matter':
the (silicon) cores are there and usable,
but (because of the thermal dissipation) the large amount of cores behaves differently.

Moreover, the parallel computing introduced the "dark performance".
Due to the classic computing principles,
the first core must speak to all fellow cores,
and \textit{this non-parallelizable fraction of the
	time increases with the
	number of the cores}~\cite{VeghRoofline:2019,VeghPerformanceBook:2019}. The result is that the top supercomputers show an efficacy around only 1\% when solving real-life tasks. Another analogy is made with the "gravitational collapse", where a "communicational collapse" is demonstrated in Fig.~5.(a) in~\cite{CommunicationCollapse:2018}: showing that at extremely large number of cores that exceeds  communication intensity leads to unexpected and drastic changes of the network latency.

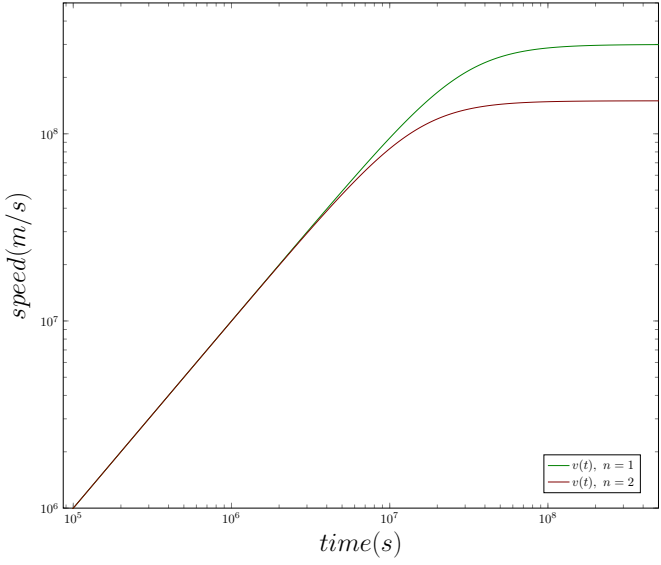
\begin{figure}
	\maxsizebox{\columnwidth}{!}
	{
		\begin{tikzpicture}%[scale=1.5]
		\begin{axis}[
		%  axis y line*=left,
		title={\huge Relativistic speed of body accelerated by 'g'},
		width=\textwidth,
		xlabel=\huge $time(s)$,
		ylabel=\huge $speed (m/s)$,
		ymin=1e6, ymax=5e8,
		xmin=\OneDay, xmax=5e8,
		xmode=log,
		log basis x=10,
		ymode=log,
		log basis y=10,
		legend style={
			cells={anchor=east},
			legend pos={south east},
		},
		]
		
		\addlegendentry{$v(t),~n=1$}
		\addplot[samples=501,domain=\OneDay:1e9,webgreen]
		{\RelSpeed} ;\label{plot_loo}\OneDay
		\addplot[samples=501,domain=\OneDay:1e9,webred]
		{\RelSpeedB} ;\label{plot_loo}
		\addlegendentry{$v(t),~n=2$}
		
		\end{axis}

		\end{tikzpicture}
	}
	\label{fig:lightspeed}
	\caption{How the speed of a body accelerated by $g$ depends on the time, in relativistic approach, see also table~\ref{tab:summing}. Compare to Figs.~4 %\ref{fig:SupercompSpeedOfLight}
		  and~6%\ref{fig:exaRMaxAlpha}
	 }
\vskip-\baselineskip
\end{figure}

The  signals propagation time is also very much similar 
to the finite propagation of the physical fields, and even the latency time of the interfaces can be paired with
creating and attenuating the physical carriers. Zero-time on/off signals are possible both in the classical physics and classical computing,
while in the corresponding modern counterparts, to create, transfer and detect signals 
time must be accounted for; the effect noticed as that the time delay through wires (in this extended sense) grows compared to the time of gating~\cite{LimitsOfLimits2014}. 

%Note that also the "black holes" have their analogs in computing.
%Adding more nominal performance (more cores) has no noticable effect: the "black hole" does not enable to emit more payload performance than a specific (environment-dependent) maximum.
%

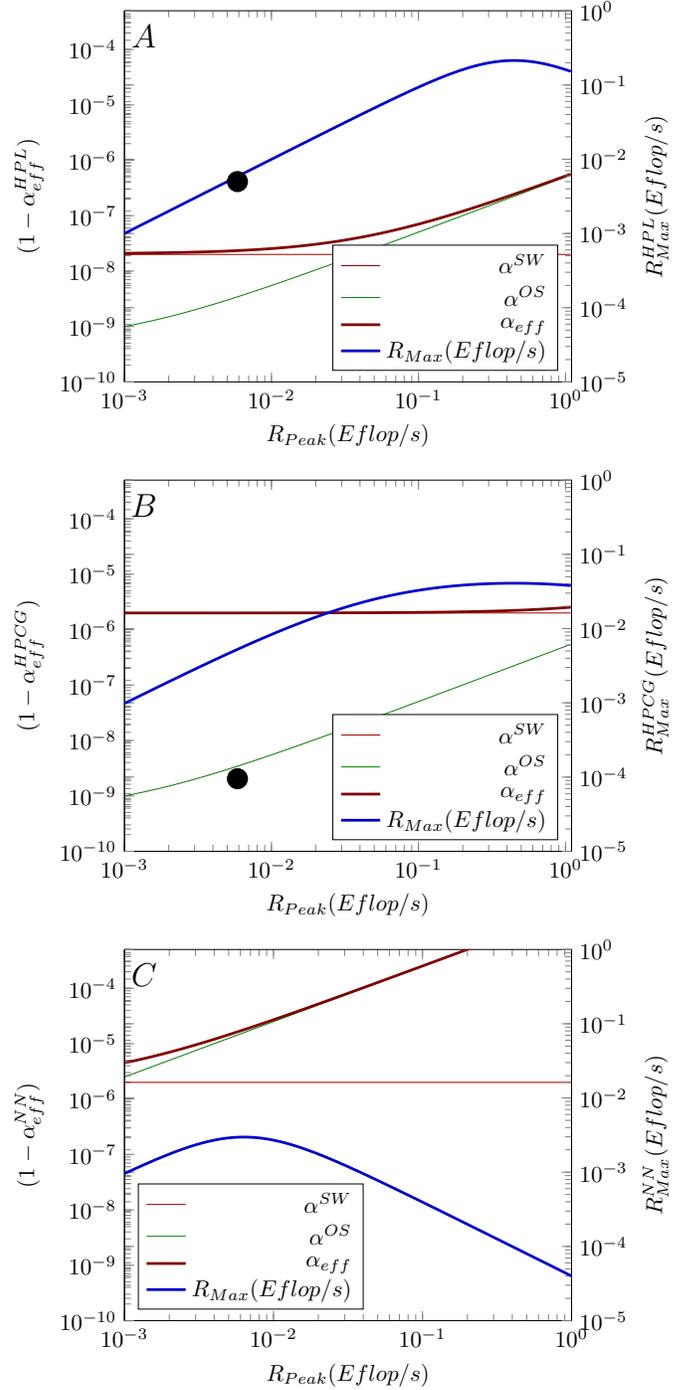
\begin{figure}
	\hskip-1.75cm\maxsizebox{1.25\columnwidth}{!}
	{
		\def\constProcFreq{1}	% processor frequency, GHz
		\def\constMPE{1} % If MPE trick (see Taihulight) in use
		\def\constProcPerformance{(100*\constMPE)}  %processor performance, GFlops
		\def\constNoOfProcessors{x*1e9/\constProcPerformance} % x in Eflop/s
		\def\constTotalClocks{2e13} % Total measurement time, in ticks
		\def\constContextChange{1e4} % Time of context change, in ticks

		%% Derived 
		
		%\def\constProcFreq{1}	% processor frequency, GHz
		\def\constMicroSecToTicks{1e3*\constProcFreq}  % usec time to clock ticks
		%\def\constNoOfProcessors{x*1e8/64}
		% Calculate different \alpha contributions
		\def\constAlphaContext{\constContextChange/\constTotalClocks}
		\def\constAlphaLoop{\constNoOfProcessors/\constTotalClocks}
		\def\constAlphaOS{\constAlphaContext
			+\constAlphaLoop
		}
		% The sum of all contributions
		\def\constAlphaTotal{(\constAlphaSW+\constAlphaOS)}
		\def\constMinusAlpha{1-\constAlphaTotal}
		\def\constEfficiency{(\constNoOfProcessors*\constAlphaTotal+\constMinusAlpha)}
		\def\constRMax{x/\constEfficiency}
		\def\constPropagationDelay{\constNoOfProcessors*1e-6/2e-8*2e9/\constTotalClocks*1000}
		
		\begin{tabular}{c}
			\begin{tikzpicture}
			
			\def\constAlphaSW{2e-8}
			
			\pgfplotsset{
				%    scale only axis,
				%    scaled x ticks=base 10:3,
				xmin=0.001, xmax=1.1,
			}
			
			\begin{axis}[
			axis y line*=left,
			legend style={
				cells={anchor=east},
				legend pos={north west},
			},
			xlabel=$R_{Peak}(Eflop/s)$,
			ylabel=$(1-\alpha_{eff}^{HPL})$,
			ymin=1e-10, ymax=5e-4,
			xmode=log,
			log basis x=10,
			ymode=log,
			log basis y=10,
			]
			% The SW contribution is constant
			\addplot[samples=501,domain=.001:1.1,webbrown]
			{\constAlphaSW }; \label{plot_SW}
			
			% Now calculate contribution of the OS
			\addplot[samples=501,domain=.001:1.1,webgreen]
			{\constAlphaOS} ;\label{plot_loop}
			
			%% Calculate propagation delay
			%\addplot[samples=501,domain=.001:1.1,webyellow]
			%{\constPropagationDelay} ;\label{plot_delay}
			%
			% Calculate total alpha 
			\addplot[samples=501,domain=.001:1.1,webred,very thick]
			{\constAlphaTotal} ;\label{plot_total}
			
			\end{axis}
			
			\begin{axis}[
			%  axis y line*=right,
			ylabel near ticks, yticklabel pos=right,
			axis x line=none,
			ymin=0.00001, ymax=1.0,
			xmode=log,
			log basis x=10,
			ymode=log,
			log basis y=10,
			ylabel=$R_{Max}^{HPL}(Eflop/s)$,
			legend style={
				cells={anchor=east},
				legend pos={south east},
			},
			]
			\addlegendimage{/pgfplots/refstyle=plot_SW}
			\addlegendentry{$\alpha^{SW}$}
			\addlegendimage{/pgfplots/refstyle=plot_loop}
			\addlegendentry{$\alpha^{OS}$}
			%\addlegendimage{/pgfplots/refstyle=plot_delay}
			%\addlegendentry{$\alpha^{delay}$}
			\addlegendimage{/pgfplots/refstyle=plot_total}
			\addlegendentry{$\alpha_{eff}$}
			%
			%%% Calculate efficiency
			%\addplot[samples=501,domain=.001:1.1,webblue] 
			%{\constRMax/x};\label{plot_eff}
			
			%% Calculate Rmax
			\addplot[samples=501,domain=.001:1.1,webblue,very thick] 
			{\constRMax};\label{plot_rmax}
			
			%\addlegendimage{/pgfplots/refstyle=plot_eff}
			%\addlegendentry{$Efficiency$}
			\addlegendimage{/pgfplots/refstyle=plot_rmax}
			\addlegendentry{$R_{Max}(Eflop/s)$}
			\addplot[only marks,  mark=*,  mark size=4, very thick] plot coordinates {
				(0.00587,0.005) %JUQUEEN computer at HPL
			};
			
			\end{axis}
			\draw node at (0.3, 5.3)  (A) {\Large $A$};
			
			\end{tikzpicture}
			\\		
			\begin{tikzpicture}
			
			\def\constAlphaSW{2e-6}
			
			\pgfplotsset{
				%    scale only axis,
				%    scaled x ticks=base 10:3,
				xmin=0.001, xmax=1.1,
			}
			
			\begin{axis}[
			axis y line*=left,
			xlabel=$R_{Peak}(Eflop/s)$,
			ylabel=$(1-\alpha_{eff}^{HPCG})$,
			ymin=1e-10, ymax=5e-4,
			xmode=log,
			log basis x=10,
			ymode=log,
			log basis y=10,
			]
			% The SW contribution is constant
			\addplot[samples=501,domain=.001:1.1,webbrown]
			{\constAlphaSW }; \label{plot_SW}
			
			% Now calculate contribution of the OS
			\addplot[samples=501,domain=.001:1.1,webgreen]
			{\constAlphaOS} ;\label{plot_loop}
			
			%% Calculate propagation delay
			%\addplot[samples=501,domain=.001:1.1,webyellow]
			%{\constPropagationDelay} ;\label{plot_delay}
			%
			% Calculate total alpha 
			\addplot[samples=501,domain=.001:1.1,webred,very thick]
			{\constAlphaTotal} ;\label{plot_total}
			
			\end{axis}
			
			\begin{axis}[
			%  axis y line*=right,
			ylabel near ticks, yticklabel pos=right,
			axis x line=none,
			ymin=0.00001, ymax=1.0,
			xmode=log,
			log basis x=10,
			ymode=log,
			log basis y=10,
			ylabel=$R_{Max}^{HPCG}(Eflop/s)$,
			legend style={
				cells={anchor=east},
				legend pos={south east},
			},
			]
			\addlegendimage{/pgfplots/refstyle=plot_SW}
			\addlegendentry{$\alpha^{SW}$}
			\addlegendimage{/pgfplots/refstyle=plot_loop}
			\addlegendentry{$\alpha^{OS}$}
			%\addlegendimage{/pgfplots/refstyle=plot_delay}
			%\addlegendentry{$\alpha^{delay}$}
			\addlegendimage{/pgfplots/refstyle=plot_total}
			\addlegendentry{$\alpha_{eff}$}
			%
			%%%% Calculate efficiency
			%\addplot[samples=501,domain=.001:1.1,webblue] 
			%{\constRMax/x};\label{plot_eff}
			
			%% Calculate Rmax
			\addplot[samples=501,domain=.001:1.1,webblue,very thick] 
			{\constRMax};\label{plot_rmax}
			
			%\addlegendimage{/pgfplots/refstyle=plot_eff}
			%\addlegendentry{$Efficiency$}
			\addlegendimage{/pgfplots/refstyle=plot_rmax}
			\addlegendentry{$R_{Max}(Eflop/s)$}
			\addplot[only marks,  mark=*,  mark size=4, very thick] plot coordinates {
				(0.00587,0.000095) %JUQUEEN computer at HPCG
			};
			
			\end{axis}
			\draw node at (0.3, 5.3)  (B) {\Large $B$};
			
			\end{tikzpicture}
			\\
			
			\begin{tikzpicture}
			
			\def\constAlphaSW{2e-6}
			\def\constAlphaLoop{5000*\constNoOfProcessors/\constTotalClocks}
			
			\pgfplotsset{
				%    scale only axis,
				%    scaled x ticks=base 10:3,
				xmin=0.001, xmax=1.0,
			}
			
			\begin{axis}[
			axis y line*=left,
			xlabel=$R_{Peak}(Eflop/s)$,
			ylabel=$(1-\alpha_{eff}^{NN})$,
			ymin=1e-10, ymax=5e-4,
			xmode=log,
			log basis x=10,
			ymode=log,
			log basis y=10,
			]
			% The SW contribution is constant
			\addplot[samples=501,domain=.001:1.1,webbrown]
			{\constAlphaSW }; \label{plot_SW}
			
			% Now calculate contribution of the OS
			\addplot[samples=501,domain=.001:1.1,webgreen]
			{\constAlphaOS} ;\label{plot_loop}
			
			%% Calculate propagation delay
			%\addplot[samples=501,domain=.001:1.1,webyellow]
			%{\constPropagationDelay} ;\label{plot_delay}
			%
			% Calculate total alpha 
			\addplot[samples=501,domain=.001:1.1,webred,very thick]
			{\constAlphaTotal} ;\label{plot_total}
			
			\end{axis}
			
			\begin{axis}[
			%  axis y line*=right,
			ylabel near ticks, yticklabel pos=right,
			axis x line=none,
			ymin=0.00001, ymax=1.0,
			xmode=log,
			log basis x=10,
			ymode=log,
			log basis y=10,
			ylabel=$R_{Max}^{NN}(Eflop/s)$,
			legend style={
				cells={anchor=east},
				legend pos={south west},
			},
			]
			\addlegendimage{/pgfplots/refstyle=plot_SW}
			\addlegendentry{$\alpha^{SW}$}
			\addlegendimage{/pgfplots/refstyle=plot_loop}
			\addlegendentry{$\alpha^{OS}$}
			%\addlegendimage{/pgfplots/refstyle=plot_delay}
			%\addlegendentry{$\alpha^{delay}$}
			\addlegendimage{/pgfplots/refstyle=plot_total}
			\addlegendentry{$\alpha_{eff}$}
			%
			%%%% Calculate efficiency
			%\addplot[samples=501,domain=.001:1.1,webblue] 
			%{\constRMax/x};\label{plot_eff}
			
			%% Calculate Rmax
			\addplot[samples=501,domain=.001:1.1,webblue,very thick] 
			{\constRMax};\label{plot_rmax}
			
			%\addlegendimage{/pgfplots/refstyle=plot_eff}
			%\addlegendentry{$Efficiency$}
			\addlegendimage{/pgfplots/refstyle=plot_rmax}
			\addlegendentry{$R_{Max}(Eflop/s)$}
			
			%		\addplot[only marks,  mark=*,  mark size=4, very thick] plot coordinates {
			%	(0.00587,0.000095) %JUQUEEN computer at HPCG
			%};
			
			\end{axis}
			\draw node at (0.3, 5.3)  (C) {\Large $C$};
			
			\end{tikzpicture}
		\end{tabular}
	}
	\caption{Contributions $(1-\alpha_{eff}^X)$ to $(1-\alpha_{eff}^{total})$ and max payload performance $R_{Max}$ of a fictive supercomputer ($P=1Gflop/s$ @ $1GHz$) in function of the nominal performance.
		The blue diagram line refers to the right hand scale ($R_{Max}$ values), all others (the $(1-\alpha_{eff}^{X})$ contributions) to the left scale. The figure is purely illustrating the concepts; the displayed numbers are somewhat similar to the real ones.  The black dots mark the \gls{HPL} and  \gls{HPCG} performance, respectively, of the computer used in works~\cite{NeuralNetworkPerformance:2018,NeuralScaling2017}.	
	}
	\label{fig:alphacontributions}	
\vskip-\baselineskip
\end{figure}

Figure~\ref{fig:alphacontributions} introduces a further analogy. The non-parallelizable fraction (denoted on the figure by $\alpha_{eff}^{X}$) of the computing task comprises components of different origin. As already discussed, and was noticed decades ago, "\textit{the inherent communication-to-computation ratio in a
	parallel application is one of the important determinants
	of its performance on any architecture}"~\cite{ScalingParallel:1993}, 
suggesting that the communication can be a dominant contribution  in system’s performance.
Figure.~\ref{fig:alphacontributions}.A displays the case of a minimum communication, and Figure~\ref{fig:alphacontributions}.B the moderately increased one
(corresponding to real-life supercomputer tasks).
As the nominal performance increases linearly and the performance decreases exponentially with the number of cores, 
at some critical value where an inflection point occurs, the
resulting performance starts to decrease.
The resulting large non-parallelizable fraction strongly decreases the efficacy
(or in other words: the performance gain or speedup) of the system~\cite{VeghPerformanceWall:2019,VeghSupercomp:2018}.
The effect was noticed early~\cite{ScalingParallel:1993}, under  different
 technical conditions but forgotten due to the successes of the development of the parallelization technology.

Figure 6.A illustrates the behavior measured with  \gls{HPL} benchmark. The looping contribution becomes remarkable around 0.1~Eflops, and breaks down payload performance when approaching 1~Eflops (see also Fig.~1 in~\cite{ScalingParallel:1993}).
In figure 6.B the behavior measured with benchmark \gls{HPCG} is displayed. In this case the contribution of the application (brown line) is much higher, the looping contribution (thin green line) is the same as above. 
Consequently, the achievable payload performance is lower and also the breakdown of the performance is softer.

\subsection{Analogy with the quantum physics}
\label{sec:AnalogyQuantum}
The electronic computers are clock-driven systems, i.e. no action can happen in a time period shorter than the length of one clock period.   The typical value of that "quantum of time" today is in the nanosecond range, so in the everyday practice, the time seems to be continuous and 
%the time difference of actions we can perceive is in the range of dozens of milliseconds, so here
the "quantal nature of time" cannot be noticed.  Some (sequential) non-payload fragment in the total time is always present in the parallelized sequential systems: it cannot be smaller than the ratio of the length of two clock periods divided by the total measurement time, since forking and joining the other threads cannot be shorter than one clock period. 
Unfortunately, the technical implementation needs about ten thousand times longer time~\cite{Tsafrir:2007,armContextSwitching:2007}. 

The total time of the performance measurement is large (typically hours) but finite, so the non-parallelizable fraction is small but finite.
As discussed, the latter \textit{increases with the number of the cores in the system}. Because of this,
in contrast with the statement in~\cite{ScalingParallel:1993} that "\textit{the serial fraction \dots is a diminishing function of the problem size}", at sufficiently large number of cores
the serial fraction even  starts to dominate.
Because of Amdahl's Law, the absolute value of \textit{the computing performance of parallelized systems
	has inherently an upper limit, and the efficiency
% (the payload performance)
is the lower the higher is the
number of the aggregated processing units}.

The processor-based brain simulation provides an
"experimental evidence"~\cite{VeghBrainAmdahl:2019} that the time in computing shows quantal behavior, analogously with the energy in physics. When simulating neurons using processors,
the ratio of the simulated (biological) time and the processor time used to simulate the biological effect may considerably differ, so to avoid working with "signals from the future", periodic synchronization is required that introduces a special "biological clock cycle".
The role of this clock period is the same as that of the clock signal in the clocked digital electronics: what happens in this period, it happens "at the same time"\footnote{This periodic synchronization will be a limiting factor in large-scale utilization of processor-based artificial neural chips~\cite{IntelLoihi:2018}, although thanks to the cca. thousand times higher "single-processor performance", only when approaching the computing capacity of (part of) the brain.}.

The brain simulation (and in somewhat smaller scale: artificial neural computing) requires intensive data exchange between the parallel threads:
the neurons are expected to tell the result of their neural calculations periodically to thousands of fellow neurons.
The commonly used $1~ms$ "grid time" is,
 however, $10^6$ times longer than the
$1~ns$ clock cycle common in the digital electronics~\cite{NeuralNetworkPerformance:2018}.
Correspondingly, its influence on the performance is noticeable.
Figure~6%\ref{fig:alphacontributions}
.C demonstrates what happens if the clock cycle is 5000 times longer than in Figure~6%\ref{fig:alphacontributions}
.B:
it causes a drastic decrease in the achievable performance and strongly shifts the performance breakdown toward lower nominal performance values.
As shown, the "quantal nature of time" in computing
changes the behavior of the performance drastically.

In addition, the thousands times more communication contributes considerably to the non-payload sequential-only fraction,
so it degrades further the efficacy of the computing system.  What is worse, they are expected to send their messages at the end of the grid time period, causing a huge burst of messages.

Not only the achievable performance is by orders of magnitude lower,
but also the "communicational collapse" (see also~\cite{ScalingParallel:1993}) occurs at orders of
magnitude lower nominal performance. 
This is the reason why less than one percent of the planned capacity
can be achieved even by the custom built large scale ANN simulators~\cite{SpiNNaker:2013}.
Similarly, the SW and HW based simulators show up the same limitation~\cite{NeuralNetworkPerformance:2018,VeghBrainAmdahl:2019}.
This is why only a few dozens of thousands of neurons can be simulated on processor-based brain simulators (including both the many-thread software simulators and the purpose-built brain simulator)~\cite{NeuralNetworkPerformance:2018}.
The memory of extremely large supercomputers can be populated with objects simulating neurons~\cite{SpikingPetascale2014}, but as soon as they need to communicate,
the task collapses as predicted in Fig.~\ref{fig:alphacontributions}. This is indirectly underpinned~\cite{NeuralScaling2017} by that the different handling
of the threads changes the efficacy sensitively and that the time required for more detailed simulation increases non-linearly~\cite{NeuralNetworkPerformance:2018,VeghBrainAmdahl:2019}.

\subsection{Analogy with the interactions of particles}
\label{sec:AnalogyInteraction}
In the ’classic computing’ the processors ability to \textit{communicating} with each other is not a native feature:
in the \textit{\textbf{S}ingle \textbf{P}rocessor \textbf{A}pproach} questions like message sending to and receiving from some other party as well as sharing resources
has no sense at all (as no other party exists); messaging is very ineffectively imitated by SW in the layer between HW and the real SW. \textit{This feature alone prevents building exascale supercomputers}~\cite{VeghPerformanceWall:2019}: 
after reaching a critical number of processors, \textit{adding more processors leads to a decreasing resulting performance}~\cite{VeghPerformanceWall:2019,VeghBrainAmdahl:2019}, as experienced in~\cite{ScalingParallel:1993,NeuralScaling2017}
and caused demonstrative failures such as the cases of $Gyoukou$ or $Aurora$ or $SpiNNaker$. This critical number (using the present technology and implementation) is under 10M cores; the only exception (as of end of 2019) is $Taihulight$,
because it is using a slightly different computing principle
with its cooperating processors~\cite{CooperativeComputing2015}.

The laws of parallel computing result in the actual behavior of the computing systems the more difference from that expected on the basis of the classical computing
the more communication takes place~\cite{VeghRoofline:2019}. Similarly, in  physics, the behavior of an atom is strongly changed by the interaction (communication) with other particles.

This phenomenon cannot be explained in the "classic computing" frame.
The limits of single-processor performance enforced by the laws of nature~\cite{LimitsOfLimits2014} are topped by the limitations
 of parallel computing~\cite{VeghPerformanceWall:2019,VeghRoofline:2019},
and further limited through introducing the "biological clock period"~\cite{VeghBrainAmdahl:2019}.
Notice that these contributions are competing with each other,
the actual circumstances decide which of them will dominate.
Their effect, however, is very similar: according to Amdahl,
\textit{what is not parallel is qualified as sequential}.

\subsection{Analogy with the uncertainty principle}
\label{sec:AnalogyUncertaintyPrinciple}

Even the quantum physical uncertainty principle which states that (unlike in classical physics)
one cannot measure 
accurately certain pairs of physical properties of a particle (like the position and the momentum) at the same time,
has its counterpart in computing.
Using registers (and caches and pipelines), one can perform computations with much higher speed, but to service an interrupt,
one has to save/restore registers and renew cache content.
Similar is the case with accelerators:
copying data from one memory to another
or dealing with coherence increases latency. That is, \textit{one cannot have low latency and high
	performance at the same time}, using the same processor design principles. 
The same processor design principles cannot be equally good for preparing
high-performance single thread applications and high performance parallelized sequential systems.

\section{The classic versus modern paradigm}\label{sec:ModernParadigm}

Today we have extremely inexpensive (and at the same time: extremely complex and 
powerful) processors around (a "free resource"~\cite{SpiNNaker:2013})
and we come to the age when no additional reasonable 
functionality can be implemented in processors through adding more transistors, the over-engineered processors optimized for single-processor regime do not enable reducing the clock period~\cite{EPIC:2000}. 
The computing power hidden in many-core processors cannot be utilized effectively 
for payload work, because of the "power wall" (\textit{partly because of the improper working regime}~\cite{EnergyProportional2007}):
we came at the age of "dark 
silicon"~\cite{Computing_Dark_Silicon_2017},
we have "too many" processors~\cite{TooManyCores2007} around.
The supercomputers face critical efficiency and performance issues; 
the real-time (especially the cyber-physical) systems experience serious predictability, 
latency and throughput issues;  in summary, the computing performance (without 
changing the present paradigm) reached its technological bounds. 
Computing needs renewal~\cite{RenewingComputingVegh:2018}.
Our proposal, the \textit{Explicitly Many-Processor Approach (EMPA)}~\cite{IntroducingEMPA2018},
is \textit{to introduce a new computing paradigm} and through that \textit{to reshape the way in which computers are designed and used} today.

\subsection{Overview of the modern paradigm}
The new paradigm is based on making fine distinctions in  specific points, which are also present in the old paradigm. Those points, however, must be scrutinized in all \textit{occurring cases}, and
whether and how long can they can be neglected.  These points are:

\begin{itemize}
	\item consider explicitly that\textit{ not only one processor} (aka Central Processing Unit) exists, i.e.
	\begin{itemize}
		\item the processing capability (akin to the data storage capability) is \textit{one of the resources} rather than a central singleton
		\item not necessarily \textit{the same processing unit} (out of the several identical ones) is used to solve all parts of the problem
		\item a kind of redundancy (an easy method of replacing a flawed processing unit) through using virtual processing units is provided (mainly \textit{to increase the mean time between technical errors}), like ~\cite{ARM:big.LITTLE:2011,Congy:CoreSpilling:2007}
		\item \textit{different processors can and must cooperate}  in solving a task,
		i.e. direct data and control exchange among the processing units are made possible; the ability to communicate with other processing units, similar to~\cite{CooperativeComputing2015}, is a native feature %of the processing units
		\item \textit{flexibility for making ad-hoc assemblies} for more efficient processing is provided
		\item the large number of processors is used for unusual tasks, such as \textit{replacing memory operation with using additional processors}
	\end{itemize}
	\item the misconception of the segregated computer components is reinterpreted
	\begin{itemize}
		\item \textit{the efficacy of utilization of the several processors
			is increased} by using multi-port memories (similar to~\cite{Cypress15}) 
		\item a "memory only" concept (somewhat similar to that in~\cite{ScratchpadMemory:2002}) is introduced (as opposed to the "registers only" concept), using purpose-oriented, optionally distributed, partly local, memory banks
		\item the principle of locality is introduced into memory handling at hardware level, through introducing hierarchic buses
	\end{itemize}
	
	\item the misconception of the "sequential only" execution~\cite{BackusNeumannProgrammingStyle} is reinterpreted
	\begin{itemize}
		\item von Neumann required only "proper sequencing" for the single processing unit; this is \textit{extended} to several processing units
		\item the tasks are broken into reasonably sized and logically interconnected fragments, unlike unreasonably fragmented by the scheduler
		\item the "one-processor-one process" principle remains valid for the a task fragments, but not necessarily for the complete task
		\item the fragments can be executed in parallel if  both data dependence and hardware availability enables it (another kind of asynchronous computing~\cite{IBMAsynchronousAPI2017})
	\end{itemize}
	\item a closer hardware/software cooperation is elaborated
	\begin{itemize}
		\item the hardware and software only exist together (akin to "stack memory")
		\item when a hardware has no duty, it can sleep ("does not exist" and does not take power)
		\item the overwhelming part of the duties of \textit{synchronization, scheduling, etc.} of the OS are taken over by the hardware
		\item the compiler helps the processor with compile-time information and the processor is able to adapt (configure) itself to the task depending on the actual hardware availability
		\item strong support for  multi-threading and resource sharing, as well as low real-time latency is provided, at processor level
		\item the internal latency of the assembled large-scale systems is much reduced, while their performance is considerably enhanced
		\item the task fragments are able to return control voluntarily without the intervention of the \gls{OS}, enabling to implement more effective and more simple operating systems
	\end{itemize}
\end{itemize}

\subsection{Details of the concept}

Our proposal introduces a new concept that permits working
with \textit{virtual processors} at programming level and their mapping to physical cores at runtime level, i.e. to \textit{let 
	the computing system to adapt itself to the task}. 
A major idea of \textit{EMPA} (for an early and less mature version see~\cite{IntroducingEMPA2018}) is to use of \textit{quasi-thread (QT)} as atomic unit of processing 
that comprises both the \textit{HW} (the physical core) and the \textit{SW} (the 
code fragment running on the core). This idea was derived with 
having in mind the best features of both the \textit{HW} core and the \textit{SW} thread.
In analogy with physics, the  \textit{QT}s have "\textit{dual nature}": in the HW world of the "classic computing"
they are represented as a 'core', in the SW world as a 'thread'. However, they are the same entity
in the sense of the 'modern computing'.
The terms 'core' and 'thread' are borrowed from the conventional computing, 
but in the 'modern computing' they can actually exist only together in a time-limited way\footnote{Akin to dynamic variables on the stack: their lifetime is limited to the period when the HW and SW are properly connected. The physical memory is always there, but it is "stack memory" only when properly handled by the HW/SW components.}. \textit{ EMPA is a new computing paradigm} which needs a new underlying architecture, \textit{rather than a  new kind of parallel processing} 
running on a conventional architecture,
so \textit{it can be reasonably compared to the terms and ideas used in conventional computing only in a very limited way};
although many of its ideas and solutions are adapted from the
'classic computing'.

%Correspondingly the size of a \textit{QT} is between one machine instruction and 
%a complete thread, 
The executable task is broken into reasonably sized and loosely dependent \textit{QT}s. (The \textit{QT}s can optionally be nested into each other, akin to subroutines.) In \textit{ EMPA} \textit{for every new QT   
	a new independent processing unit (PU) is also implied, the internals (PC and part of registers) are set up properly},
and they execute their task independently\footnote{Although the idea of executing the single-thread task "in pieces" may look strange for the first moment,
	actually the same happens when the OS schedules/blocks a task.
	The key differences are that in EMPA \textit{not the same} processor is used, the \textit{QT}s are cut into fragments in a  reasonable way (preventing issues like priority inversion~\cite{PriorityInversion:1993}), the \textit{QT}s can be processed \textit{at the same time} as long as their mathematical dependence \underline{and} the actual HW resource availability enable it.}
(but under the supervision of the processor
comprising the cores).

In other words: \textit{the processing capacity is considered 
	as a resource} in the same sense as the memory is considered as a storage resource. This approach enables the 
programmer to work with 
virtual processors (mapped to physical \textit{PU}s by the computer at run-time)
and they can utilize the quick resource \textit{PU}s where they can replace utilizing 
the slow resource memory (say, hiring a quick processor from a core pool can be competitive with saving and restoring registers in the slow memory, for example when making a recursive call). The third major idea is that \textit{the PUs can cooperate} 
in various ways, including data and control synchronization, as well as \textit{outsourcing part of the received job (received as a nested \textit{QT})} to a helper core. 
An obvious example is to outsource the housekeeping activity of loop organization to a helper core: counting, addressing, comparing, etc. can be done by a helper core, while the main calculation remains to the originally delegated core. As the mapping  to physical cores occurs at runtime, (depending on the actual HW availability)
the processor can eliminate the (maybe temporarily) denied cores as well as to adapt the resource need (requested by the compiler)  of the task to the actual computing resource availability.

The processor has an additional control layer for organizing the joint work of its cores. The
cores have just a few extra communication signals and are able to execute both
conventional and so called meta-instructions (for configuring the architecture).
The latter ones are executed 
in a co-processor style: when finding a meta-instruction, the core notifies 
the processor which suspends the conventional operation of the core, controls executing
the meta-instruction (utilizing the resources of the core, providing helper cores and handling the connections between the cores as requested) then resumes core operation.

The processor needs to find the needed \textit{PUs} (cores) and the processing 
ability has to accommodate to the task; quickly, flexibly, 
effectively and inexpensively. \textit{A kind of `On demand' computing that works 
	`As-a-Service'}. This is a task not only for the processor but the complete computing 
system must participate and for that goal the complete computing stack must be 
rebuilt. 

Behind the former attempts to optimize code execution inside the processor there was no established theory, and they actually could achieve only moderate success  
because in \textit{SPA} the processor is working in real time, it has not enough resources, 
knowledge and time do discover those options completely~\cite{WallLimitsOfILP:1993}. 
In the classic computing, the compiler can find out anything about enhancing the performance but has no information about the actual run-time HW availability, furthermore it has no way to tell its findings to the processor. The processor has the HW availability information, but has to "reinvent the wheel" with respect to enhancing performance; in real time. In EMPA, the compiler puts its findings in the executable code in form of meta-instructions
("configware"), and the actual core executes them with the assistance of the new control layer of the processor.
The processor can choose from those 
options, considering the actual HW availability, in a style '\textbf{if} NeededNumberOfResourcesAvailable \textbf{then} Method1 \textbf{else} Method2', maybe nested one to another.

\subsection{Some advantages of EMPA}
The approach results in several considerable advantages,
but the page limit forces us to mention just a few.
\begin{itemize}
	\item as a new \textit{QT} receives a new \gls{PU}(s), there is no need to save/restore registers and return address
	(less memory utilization and less instruction cycles)
	\item the OS can receive its own  \gls{PU}, which is initialized 
	in kernel mode and can promptly (i.e. without the need of context change) service the requests from the requestor core
	\item for resource sharing, temporarily a \gls{PU} can be 
	delegated to protect the critical section; the next call to
	run the code fragment with the same offset will be delayed until the processing by the first \gls{PU} terminates
	\item the processor can natively accommodate to the variable
	need of parallelization
	\item the actually out-of-use cores are waiting in low energy consumption mode
	\item the hierarchic core-to-core communication greatly increases the memory throughput 
	\item the asynchronous-style computing~\cite{IBMasynhronousParallel:2019} largely reduces the loss due to the gap~\cite{NinjaPerformanceGap:2015:CACM} between speed of the processor and that of the memory
	\item the direct core-to-core connection (more dynamic than in~\cite{CooperativeComputing2015}) greatly enhances efficacy in large systems~\cite{TaihulightHPCG:2018}
	\item the thread-like feature to \textit{fork()} and the hierarchic buses change the dependence of on the number of cores from linear to logarithmic~\cite{VeghPerformanceWall:2019}
	(enables to build really exa-scale supercomputers)
\end{itemize}

The very first version of \textit{EMPA}~\cite{RenewingComputingVegh:2018}
has been implemented in a form of simple (practically untimed)
simulator~\cite{VeghEMPAthY86:2016}, now an advanced (Transaction Level Modelled) simulator is prepared in SystemC. The initial version adapted
Y86 cores~\cite{hallaron}, the new one RISC-V cores. %~\cite{RISCVarchitecture:2017}.
Also part-solutions are modeled in FPGA.

\section*{Summary}
The today's computing is more and more typically utilized under extreme conditions: 
 %measuring the speed of light using system having components slower than the speed of light;
providing extreme low latency time in interrupt-driven system,
extremely large computing performance in parallelly working systems, 
relatively high performance on a complex computer system for a long time, service requests in an energy aware mode.
To some measure, these activities are solved under the umbrella of the old paradigm for non-extreme scale systems.
Those experiences, however, must be reviewed when working with
extreme-large systems, because \textit{the scaling is as  nonlinear
as the phenomena are experienced}.
Consequently, scrutinizing the details of the basic principles of computing,
a "modern computing paradigm" that is able to explain the new extreme-condition phenomena on one side
and enables to build computing systems with much more advantageous features on the other side, can be constructed.

%\section{Bibliography}

\bibliographystyle{IEEEtran} 
%\bibliography{../../CommonComputerBibliography,%
%	../../CommonParallelBibliography,%
%	../../CommonNervousBibliography,%
%	../../CommonPrivateBibliography,%
%	../../CommonOtherBibliography%			
%}
\bibliography{../../CommonBibliography,%
	../../CommonPrivateBibliography%
}
\end{document}